\documentclass[fleqn,usenatbib]{mnras}

\usepackage[T1]{fontenc}

\DeclareRobustCommand{\VAN}[3]{#2}
\let\VANthebibliography\thebibliography
\def\thebibliography{\DeclareRobustCommand{\VAN}[3]{##3}\VANthebibliography}

\usepackage{graphicx}	
\usepackage{amsmath}	
\usepackage{amssymb}	
\usepackage{epstopdf}
\usepackage{soul}
\usepackage{newtxtext,newtxmath}

\title[Flare activity and metallicity]{Further evidence of the link between activity and metallicity using the flaring properties of stars in the Kepler field}

\author[V. See et al.]{
Victor See$^{1}$\thanks{E-mail: victor.see@esa.int}\thanks{ESA research fellow},
Julia Roquette$^{2}$,
Louis Amard$^{3}$,
Sean Matt$^{4}$
\\
$^{1}$European Space Agency (ESA), European Space Research and Technology Centre (ESTEC), Keplerlaan 1, 2201 AZ Noordwijk, The Netherlands\\
$^{2}$D\'{e}partement d'Astronomie, Universit\'{e} de Gen\`{e}ve, Chemin Pegasi 51, 1290 Versoix, Switzerland\\
$^{3}$D\'{e}partement d’Astrophysique/AIM, CEA/IRFU, CNRS/INSU, Univ.  Paris-Saclay \& Univ.  de Paris, 91191 Gif-sur-Yvette, France\\
$^{4}$ University of Exeter, Department of Physics \& Astronomy, Stocker Road, Devon, Exeter, EX4 4QL, UK
}

\date{Accepted XXX. Received YYY; in original form ZZZ}

\pubyear{2022}

\begin{document}
\label{firstpage}
\pagerange{\pageref{firstpage}--\pageref{lastpage}}
\maketitle

\begin{abstract}
The magnetic activity level of low-mass stars is known to vary as a function of the physical properties of the star. Many studies have shown that the stellar mass and rotation are both important parameters that determine magnetic activity levels. In contrast, the impact of a star's chemical composition on magnetic activity has received comparatively little attention. Data sets for traditional activity proxies, e.g. X-ray emission or calcium emission, are not large enough to search for metallicity trends in a statistically meaningful way. Recently, studies have used the photometric variability amplitude as a proxy for magnetic activity to investigate the role of metallicity because it can be relatively easily measured for large samples of stars. These studies find that magnetic activity and metallicity are positively correlated. In this work, we investigate the link between activity and metallicity further by studying the flaring properties of stars in the Kepler field. Similar to the photometric variability, we find that flaring activity is stronger in more metal-rich stars for a fixed mass and rotation period. This result adds to a growing body of evidence that magnetic field generation is correlated with metallicity.
\end{abstract}




\section{Introduction}
\label{sec:Intro}
Understanding the processes that govern magnetic field generation in low-mass stars ($M_\star \lesssim 1.3M_\odot$) is an ongoing task. One way to probe the magnetic field generation process is to study how the magnetic properties of low-mass stars scale with their physical properties. The most relevant parameter appears to be the Rossby number which is defined here as the rotation period of the star divided by its convective turnover time. This parameter encapsulates the interplay between rotation and convection that is thought to power the dynamo process in low-mass stars \citep[e.g.][]{Brun2017}. Numerous studies have shown that magnetism and activity is generally stronger in stars with smaller Rossby numbers up to a saturation value \citep[e.g.][]{Noyes1984,Saar1999,Pizzolato2003,Mamajek2008,Reiners2009,Vidotto2014ZDI,See2015,See2019ZB,See2019MDot,Stelzer2016,Newton2017,Wright2018,Kochukhov2020,Boudreaux2022,Reiners2022}. 

Although the Rossby number is the most relevant parameter when it comes to predicting the activity level of low-mass stars, it is also hard to estimate. The difficulty arises because the convective turnover time is not a directly observable property and is hard to constrain \citep[e.g.][]{Corsaro2021}. Therefore, it is also useful to study how magnetic activity scales with more directly measurable stellar properties such as rotation or mass. In general, more rapidly rotating stars are more magnetically active than slowly rotating stars \citep[e.g.][]{Vidotto2014ZDI}. This is consistent with the fact that stars with smaller Rossby numbers are generally more magnetically active since Rossby number is proportional to the rotation period. Additionally, less massive stars are generally more magnetically active than more massive stars. This is also consistent with low Rossby number stars having high activity because less massive stars tend to have longer convective turnover times and, therefore, smaller Rossby numbers.

While the impact of stellar mass and rotation on the activity levels of low-mass stars are well known, the impact of metallicity has received comparatively little attention until recently. Stellar structure models show that more metal-rich stars have longer convective turnover times and, therefore, smaller Rossby numbers \citep{vanSaders2013,Karoff2018,Amard2019,Amard2020RotEvo}. Therefore, one should expect that more metal-rich stars should have more efficient dynamos and be more magnetically active. However, testing this hypothesis is difficult since relatively large sample sizes are needed to properly disentangle the impact of metallicity from mass and rotation on activity levels. 

In recent years, a number of authors have investigated the link between activity and metallicity using the photometric variability amplitude as a proxy for magnetic activity \citep[e.g.][]{Karoff2018,Reinhold2020}. The advantage of using the photometric variability amplitude over more traditional proxies such as X-ray emission or calcium emission is that it can be easily estimated for large samples of stars thanks to missions like Kepler \citep[e.g.][]{Basri2010,Reinhold2013,McQuillan2014}. These investigations find that more metal-rich stars generally have larger variability amplitudes and are, therefore, more magnetically active which is in line with the theoretical expectation. Most recently, in\citet{See2021}, we studied a sample of over 3000 low-mass stars in the Kepler field covering a wide range of masses and rotation periods. Similar to previous works, we found that, at fixed mass and rotation, more metal-rich stars generally have larger photometric variability amplitudes. 

Although these studies have advanced our understanding of the role that metallicity plays in magnetic field generation, they suffer from the fact the photometric variability amplitude is a relatively indirect tracer of magnetic activity. For example, our analysis in \citet{See2021} is slightly hampered by the presence of a dip seen in the photometric variability versus Rossby number diagram that is not seen in the activity-rotation relations for other activity proxies \citep[see also][]{Reinhold2019}. Additional factors, e.g. stellar inclination \citep{Shapiro2016}, can also impact the variability amplitude and could add significant scatter to the trends being studied. Lastly, metallicity can affect the contrast of magnetic features \citep{Witzke2018,Witzke2020} and may therefore influence the photometric variability of a star in a way that is unrelated to magnetic field generation. For these reasons, it would be beneficial to investigate how metallicity affects magnetic activity using other activity proxies.

In this study, we build on our work from \citet{See2021} to investigate how the flaring properties of stars in the Kepler field depend on stellar metallicity. Flare events involve a rapid conversion of magnetic energy to electromagnetic radiation in the atmospheres of low-mass stars \citep{Benz2010,Shibata2016}. These events show up in photometric light curves as a rapid rise phase followed by an exponential decay and have been detected on wide range of stars from G dwarfs \citep{Maehara2012,Shibayama2013,Doyle2020} to K dwarfs \citep{Zaleski2022} and M dwarfs \citep{Hawley2014,Chang2017}. Due to the magnetic origin of flares, studying flaring properties, such as flare rates, flare energies or flare frequency distributions, allows us to learn about the stellar magnetic field generation process. Previous works have already shown that flaring properties vary with stellar properties like rotation, effective temperature or Rossby number in a similar way to other activity proxies \citep{Candelaresi2014,Davenport2016,Stelzer2016,Yang2019,Davenport2019,Notsu2019,Guenther2020}. Additionally, using flaring properties as an activity proxy is complementary to using the photometric variability as an activity proxy since flares are not affected by some of the previously mentioned issues that the photometric variability suffers from.

The rest of this paper is structured as follows. In section \ref{sec:Sample}, we present the sample of Kepler field stars that we use in this study. In section \ref{sec:Results}, we show how the flaring properties of this sample vary as a function of stellar properties, focussing on the Rossby number and metallicity. Finally, we present our conclusions and discuss the implications of these results in section \ref{sec:Conclusions}.

\section{Stellar Sample}
\label{sec:Sample}
The sample of stars we use for this study is an updated version of the samples used by \citet{Amard2020Kepler} and \citet{See2021}. Similar to those studies, the sample in this work is the result of cross-matching the samples from a number of different surveys and studies and focusses on stars in the Kepler field (Data Release \verb|Q1-Q17-DR25| retrieved through the NASA Exoplanet Archive\footnote{\url{https://exoplanetarchive.ipac.caltech.edu/}}). The rotation periods, $P_{\rm rot}$, are taken from either \citet{McQuillan2014} or the series of papers by \citet{Santos2019} and \citet{Santos2021}. When periods exist for a star in multiple works, we preferentially use the one from \citet{McQuillan2014} although we note that our results are not significantly different if we were to adopt the periods from \citet{Santos2019} and \citet{Santos2021} instead in these cases. Spectroscopically derived stellar parameters  (metallicities, [Fe/H], effective temperatures, $T_{\rm eff}$, and surface gravities, $\log{g}$) are taken from the APOGEE DR17 \citep{Abdurrouf2022_APOGEEdr17} and LAMOST DR7 \verb|v2| \citep{Luo2015_LAMOST,Liu2020_LAMOST_MSR,Du2021_LAMOST_DR6_7} surveys. For LAMOST DR7 the information could be from either the low resolution spectra (LRS) or medium resolution spectra (MRS) surveys. Where objects exist in multiple of these surveys, we adopted the spectral information from the survey with the highest resolution, i.e. APOGEE ($R{\sim}22,500$), followed by LAMOST MRS ($R{\sim}7,500$), and finally LAMOST LRS ($R{\sim}1,800$). For this work, we only include stars with a reported [Fe/H] uncertainty smaller than 0.1 dex. We also only include stars with effective temperatures, $T_{\rm eff}<6500{\rm K}$, since the convective regions of hotter stars become vanishingly thin. As such, the magnetic properties of these stars do not appear to follow the same trends as cooler stars \citep[e.g.][]{Kraft1967}.

Photometric and astrometric data from Gaia-DR3 were retrieved from the Gaia-Archive$@$ESA\footnote{\url{https://gea.esac.esa.int/archive/}} and cross-matched to the Kepler database using \verb|topcat| \citep{Taylor2005}. Following the recommendations in the Gaia DR3 release papers we performed the following corrections to the data. (i) We used the new $C*$ metrics defined by \citet{Riello2021} to correct for inconsistency between different passbands. (ii) We limited the effects of brightness excess towards the fainter end of the $G_\mathrm{BP}$ passband by limiting our dataset to stars brighter than $G_\mathrm{BP}=20.9$ mag \citep{Riello2021,Fabricius2021} (iii) We applied saturation corrections for the brightest stars \citep[][appendix C.1]{Riello2021}. To select the highest quality data, we only used data with \verb|visibility_periods_used|$>\!10$ and with \verb|parallax_over_error|$\geq\!10$. Finally we also limited the dataset to sources with photometric uncertainty better than 1$\%$. 

Similar to \citet{See2021}, stellar masses, $M_\star$, and convective turnover times, $\tau$, for our sample are estimated using a grid of stellar structure models from \citet{Amard2019} and an adapted maximum-likelihood interpolation tool \citep{Valle2014}. For each star, prior spectroscopic information about its metallicity and effective temperature along with absolute magnitudes from Gaia DR3 photometry are incorporated into the mass and turnover time estimates. The turnover time is estimated at half a pressure-scale height above the base of the convective zone using a mixing length theory prescription (see \citet{Charbonnel2017} for a comparison of turnover timescales at other depths).

As well as the physical properties of the stars in our sample, we also require information about their magnetic activity. In this work we focus primarily on the normalised flaring luminosity, defined as the flaring luminosity divided by the bolometric luminosity, $R_{\rm flare}=L_{\rm flare}/L_{\rm bol}$, as calculated by \citet{Yang2019} for Kepler field stars. The normalised flaring luminosity is calculated by summing up the energies of all the flares present in a photometric light curve and normalising by the bolometric luminosity energy output over the duration of the light curve (see \citet{Yang2017} and \cite{Yang2019} for further details). \citet{Yang2019} also investigated other flare properties, such as the flare frequency distribution. However, we choose to focus on the normalised flaring luminosity in this work as it is an indication of the fraction of a star's energy output that is released through flares and therefore a useful probe of the underlying dynamo. In order to calculate $R_{\rm flare}$, the bolometric luminosity is needed. In their work, \citet{Yang2019} used the KIC effective temperature to determine the bolometric luminosity, $L_{\rm bol}=4\pi r_\star^2 \sigma T_{\rm eff}^4$. In our work, we recalculate $R_{\rm flare}$ using the spectroscopically determined effective temperatures from the APOGEE and LAMOST surveys as these are more accurate. Additionally, we also use the photometric variability amplitude, $R_{\rm per}$, as calculated by \citet{McQuillan2014} in this work. 

\begin{figure}
	\includegraphics[trim=0mm 10mm 5mm 0mm,width=\columnwidth]{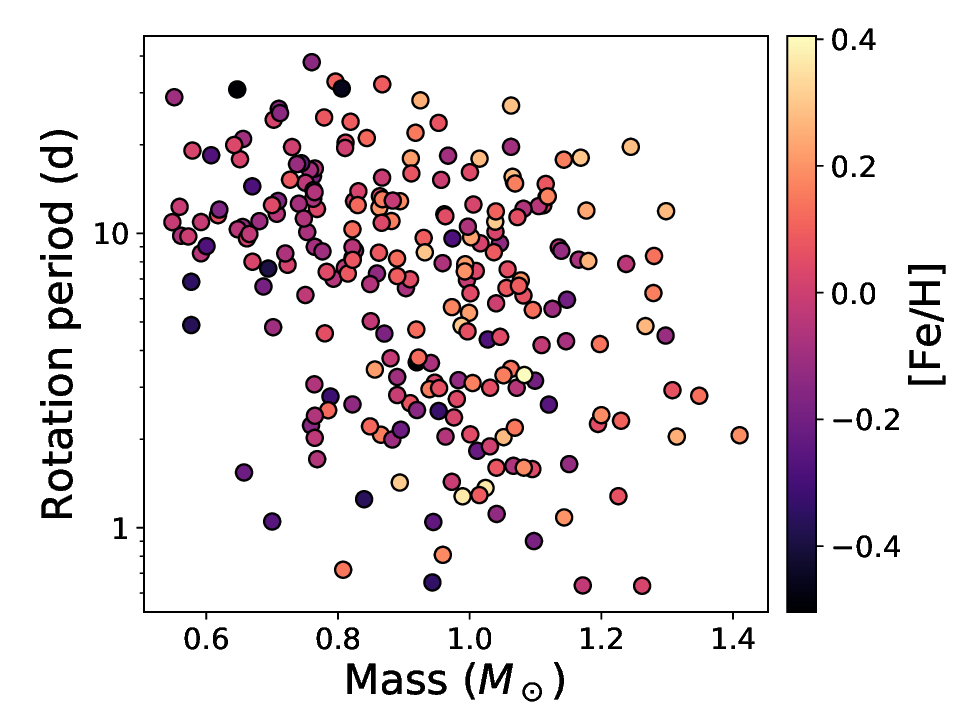}
    \caption{The masses, rotation periods and metallicities of the stars in the sample used in this work.}
    \label{fig:Properties}
\end{figure}

As we wish to focus on single main sequence stars, we removed possible near-equal-mass binaries from our sample, which typically appear as a sequence of stars 2.5$\log$(2) = 0.753 mag above the main sequence in a colour-magnitude diagram. We followed a metallicity-dependent approach, similar to the method used in \citet{Amard2020Kepler}, but with an improvement that is described in Appendix \ref{app:EqualMassBinary}, which allows us to account for the typical extinction as a function of distance in the Kepler Field. After removing the possible equal-mass binaries, we also removed sources in common with the Kepler Eclipsing Binary Catalogue \citep[V3;][]{Matijevic2012}\footnote{\url{http://keplerebs.villanova.edu/}}. Finally, we kept only sources with Gaia DR3 renormalized unit weight error \verb|ruwe|<1.4 \citep{Lindegren2018}, which selects well-behaved astrometric solutions of single stars. After these cuts, our sample consists of 240 stars. The range of masses, periods and metallicities present in our sample can be seen in fig. \ref{fig:Properties} and the numerical values of all the parameters of our stellar sample can be found in table \ref{tab:SampleParams}.

\begin{table}
	\centering
	\caption{Parameters of interest in this work for our stellar sample. The full table can be found online in a machine readable format.}
	\label{tab:SampleParams}
	\begin{tabular}{lcccccc}
		\hline
		$\rm KIC$ & $M_\star$ & $P_{\rm rot}$ & $\rm [Fe/H]$ & $\tau$ & $R_{\rm flare}$ & $R_{\rm per}$\\
		& $(M_\odot)$ & (days) & (dex) & (days) & & (ppm)\\
		\hline
1161620 & 1.07 & 6.64 & 0.09 & 19.45 & 2.91e-06 & 1.12e+04 \\
2164031 & 1.04 & 1.11 & -0.13 & 13.44 & 1.68e-06 & 1.50e+03 \\
2298664 & 0.74 & 17.31 & -0.21 & 40.00 & 1.53e-06 & 1.86e+04 \\
2441470 & 1.19 & 2.25 & 0.05 & 7.32 & 3.76e-07 & 6.50e+03 \\
2443037 & 0.71 & 11.65 & -0.04 & 53.72 & 1.40e-06 & 8.46e+03 \\
		\hline
	\end{tabular}
\end{table}

\section{Results}
\label{sec:Results}
\begin{figure}
	\includegraphics[trim=0mm 10mm 5mm 0mm,width=\columnwidth]{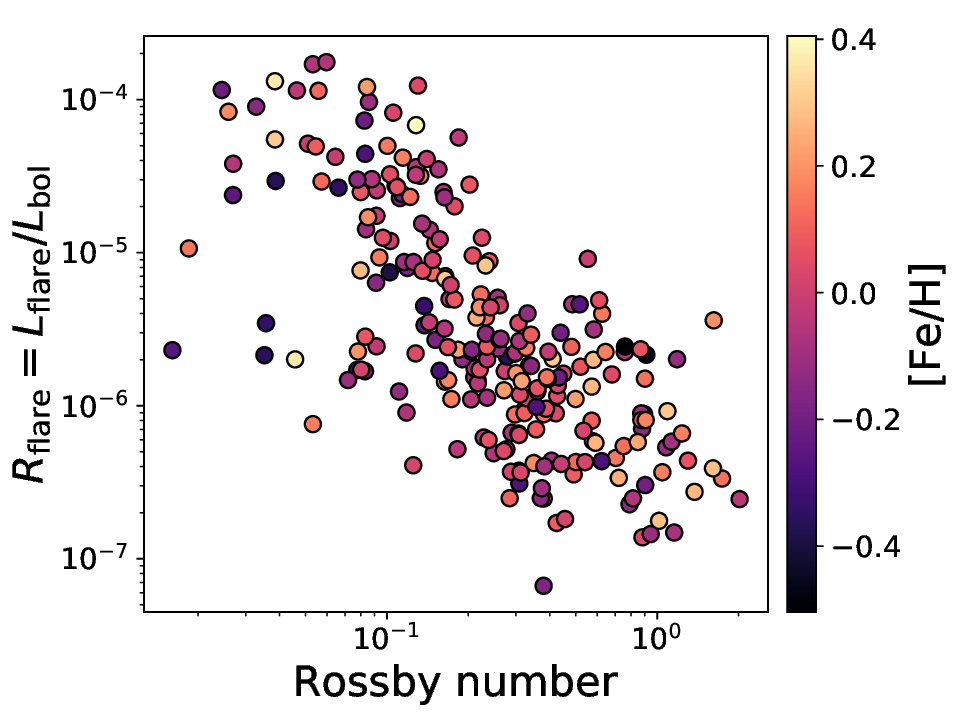}
    \caption{Normalised flare luminosity as a function of Rossby number. Each point is colour coded by metallicity, [Fe/H].}
    \label{fig:Rossby}
\end{figure}

Figure \ref{fig:Rossby} shows the normalised flare luminosity, $R_{\rm flare}$, versus Rossby number for our sample of stars. This quantity is analogous to the X-ray luminosity to bolometric luminosity ratio, $R_{\rm X}$, that is commonly used in X-ray activity studies \citep[e.g.][]{Wright2018}. Although there is a reasonable amount of scatter in fig. \ref{fig:Rossby}, an inverse power law relationship between $R_{\rm flare}$ and Rossby number is evident. This behaviour is similar to that of other activity indicators in the unsaturated regime. \citet{Yang2019} also found a similar relationship between $R_{\rm flare}$ and Rossby number for K and M type stars in the unsaturated regime. Interestingly, these authors find a much larger scatter in the $R_{\rm flare}$ vs Rossby number diagram for F and G type stars and it is not clear if these stars follow the same trends. We note that these authors do not explicitly account for any metallicity dependence when calculating the convective turnover times used in their work. 

\begin{figure*}
	\includegraphics[trim=0mm 10mm 5mm 5mm,width=\textwidth]{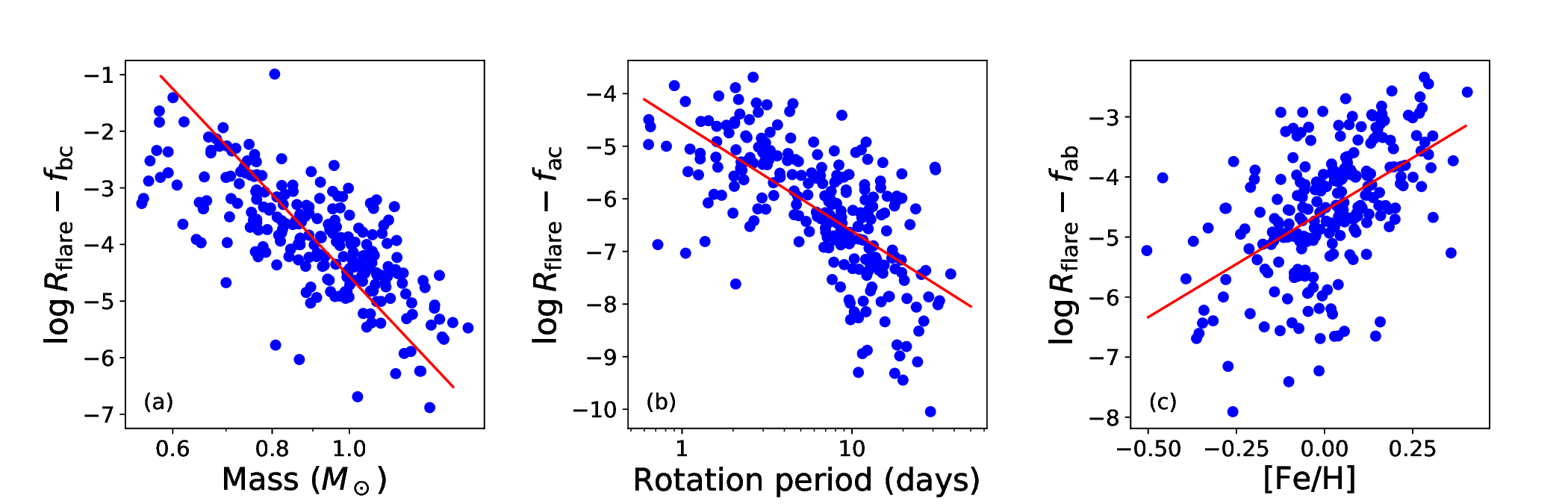}
    \caption{The relationship between the normalised flare luminosity and mass, rotation and metallicity for our sample of stars. On the y axis of each panel, the two terms on the right hand side of our multivariate fit, equation (\ref{eq:Fit}), not under consideration are subtracted from the normalised flare luminosity, i.e. $f_{\rm bc}=b\log P_{\rm rot} + c {\rm [Fe/H]}$, $f_{\rm ac}=a\log M_\star + c{\rm [Fe/H]}$ and $f_{\rm ab} = a\log M_\star + b\log P_{\rm rot}$, where $a$, $b$ and $c$ have the values given in table \ref{tab:fitParams}. The best fit lines derived from our multivariate fit are shown in red. Data points at (0.65, 0.48) in panel (a) and at (-0.86, -5.39) in panel (c) are not shown for clarity.}
    \label{fig:MultivarFit}
\end{figure*}

In \citet{See2021}, we studied how the photometric variability depends on Rossby number (see fig. 3 from that work). One surprising trend we observed is that more metal-rich stars seem to be more active even at a fixed Rossby number. Such a trend is not expected if the influence of metallicity on magnetic activity is solely through its influence on the stellar structure and, hence, the stellar dynamo. In \citet{See2021}, we suggested that this trend could be due to additional impacts that metallicity has on photometric variability that are unrelated to the dynamo, e.g. the impact of metallicity on the contrast of magnetic features at the stellar surface \citep{Witzke2018,Witzke2020}. There does not seem to be a similar residual metallicity dependence in fig. \ref{fig:Rossby}. This suggests that the residual metallicity dependence seen in fig. 3 of \citet{See2021} could be attributed to an effect that is unique to the photometric variability amplitude rather than something that is common to all activity proxies. However, we caution that our sample in this work is much smaller than our sample from \citet{See2021} and that a future study involving a larger sample could reveal a similar residual metallicity dependence in the normalised flare luminosity diagram as the one seen for the photometric variability.

In order to study the relationship between $R_{\rm flare}$ and metallicity, we would, ideally, perform a similar analysis to the one we conducted in \citet{See2021}. In that work, we divided our sample into bins of approximately constant mass and constant rotation period. This allowed us to study how magnetic activity depends on metallicity independently of the effects of mass and rotation. However, this method is not feasible for our current study due to the smaller sample size and we must take a slightly different approach. Instead, we perform an orthogonal distance multivariate regression to our full sample of the form

\begin{table}
	\centering
	\caption{Values of the fit parameters in equation (\ref{eq:Fit}) obtained from our regression.}
	\label{tab:fitParams}
	\begin{tabular}{cccc} 
		\hline
		$a$ & $b$ & $c$ & $d$\\
		\hline
		-14.95$\pm$1.36 & -2.05$\pm$0.19 & 3.54$\pm$0.49 & -4.57$\pm$0.14\\
		\hline
	\end{tabular}
\end{table}

\begin{equation}
	\log R_{\rm flare} = a \log M_\star + b \log P_{\rm rot} + c{\rm [Fe/H]} + d,
	\label{eq:Fit}
\end{equation}
where $M_\star$ is the stellar mass, $P_{\rm rot}$ is the rotation period, $\rm [Fe/H]$ is the metallicity and $a$, $b$, $c$ \& $d$ are the fit parameters. The values of these fit parameters from our regression are shown in table \ref{tab:fitParams}. This is similar to the analysis conducted by \citet{Reinhold2020} on variability data in their supplementary materials. However, we have parameterised our multivariate fit in terms of stellar mass rather than effective temperature since mass and metallicity are independent variables whereas effective temperature and metallicity are not.

\begin{figure}
	\includegraphics[trim=0mm 10mm 5mm 5mm,width=\columnwidth]{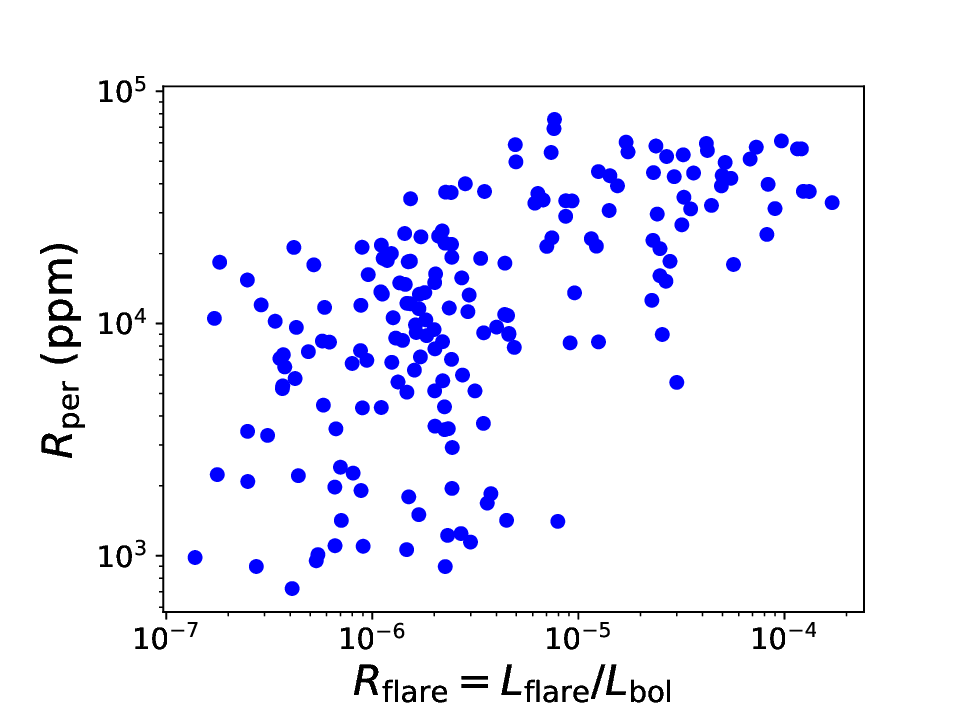}
    \caption{Photometric variability amplitude vs normalised flare luminosity.}
    \label{fig:VarVsFlare}
\end{figure}

Figure \ref{fig:MultivarFit} visually shows the results of our multivariate regression. Each panel shows how the normalised flare luminosity, $R_{\rm flare}$, of our sample varies as a function of either mass, rotation period or metallicity. The remaining two parameters that are not under consideration in each panel are subtracted from the normalised flare luminosity on the y-axis. The values of the fit parameters for the mass term, $a$, and rotation term, $b$, in equation (\ref{eq:Fit}) are both negative indicating that rapidly rotating and low-mass stars are more flare active than slowly rotating and high-mass stars. This can be seen in the left two panels of fig. \ref{fig:MultivarFit} and is also consistent with the behaviour of many other activity indicators as discussed in the introduction. The value of the fit parameter for the metallicity term, $c$, in equation (\ref{eq:Fit}) is positive indicating that metal-rich stars are more flare active than metal-poor stars. This can be seen in fig. \ref{fig:MultivarFit}c and is consistent with out results in \citet{See2021} that more metal-rich stars are generally more magnetically active.

Finally, as a direct comparison of our work from \citet{See2021} to this work, we plot the normalised flare activity, $R_{\rm flare}$, versus the photometric variability amplitude, $R_{\rm per}$, as measured by \citet{McQuillan2014} in fig. \ref{fig:VarVsFlare}. We see that the two activity indicators are correlated, which is consistent with the result of \citet{Yang2017}, although there is a large amount of scatter. This scatter is likely caused by the fact that both the flaring activity and photometric variability are relatively indirect proxies of magnetic activity. There are also the non-activity related factors mentioned in the introduction that can contribute towards the variability of a star that likely also add extra scatter to this plot (see also the discussion in section 3.3 of \citet{Yang2017} regarding the scatter in this plot).

\section{Conclusions}
\label{sec:Conclusions}
In this work, we study the flaring properties of a sample of 240 main sequence stars in the Kepler field. In particular, we investigated the dependence of the normalised flaring luminosity on stellar metallicity. For each star, we compile literature values for the rotation period \citep{McQuillan2014,Santos2019, Santos2021}, metallitiy \citep[LAMOST DR7 and APOGEE DR17:][]{Liu2020_LAMOST_MSR,Du2021_LAMOST_DR6_7,Abdurrouf2022_APOGEEdr17}, Gaia DR3 astrometry and photometry, and normalised flaring luminosity \citep{Yang2019}. Additionally, we calculate stellar masses and convective turnover times using the structure models of \citet{Amard2019}. Our sample predominantly lies in the unsaturated regime of the activity-rotation relation. Similar to previous works, e.g. \citet{Yang2019}, the normalised flaring luminosity of our sample is inversely correlated with Rossby number. We also demonstrate that metal-rich stars generally have larger normalised flaring luminosities than metal-poor stars. 

The result that more metal-rich stars have stronger flaring activity is consistent with the theoretical expectation. More metal-rich stars are expected to have longer convective turnover times resulting in smaller Rossby numbers and, therefore, should have stronger magnetic activity. Indeed, our study adds to the growing body of evidence that most, if not all, forms of magnetic activity scale with metallicity. For instance, previous studies have shown that another activity proxy, the photometric variability, is also correlated with metallicity \citep{Karoff2018,Reinhold2020,See2021}. Additionally, \citet{Amard2020Kepler} showed that more metal-rich stars in the Kepler field are, on average, spinning more slowly than metal-poor stars. They interpreted this as evidence that more metal-rich stars have stronger magnetised winds than metal-poor stars and, therefore, lose angular momentum more rapidly resulting in slower rotation at late ages \citep{Amard2020RotEvo}.

\section*{Acknowledgements}
We thank the anonymous referee for their time refereeing our manuscript. We also thank Oliver Hall for useful discussions. V.S. acknowledges support from the European Space Agency (ESA) as an ESA Research Fellow. J.R. acknowledges funding from the European Union’s Horizon 2020 research and innovation program (grant agreement No.101004141, NEMESIS).  L.A. acknowledges support from the Centre National des Etudes Spatialees (CNES) through the PLATO/AIM grant. S.M. acknowledges funding from the European Research Council (ERC), under the European Union's Horizon 2020 research and innovation program (grant agreement No. 682393 AWESoMeStars).

\emph{Software:} \texttt{matplotlib} \citep{Hunter2007}, \texttt{numpy} \citep{Harris2020}, \texttt{scipy} \citep{Virtanen2020}, \texttt{TOPCAT} \citep{Taylor2005}

\section*{Data Availability}
The data used throughout this work, i.e. the data contained in table \ref{tab:SampleParams}, will be made available via VizieR upon publication.



\bibliographystyle{mnras}
\bibliography{FlareMetal} 



\appendix
\section{Selecting single main sequence stars in the Kepler field}
\label{app:EqualMassBinary}
To reduce the number of equal-mass binaries in our dataset, we use an improved approach to the one we used in \citet{Amard2020Kepler}. First, we binned the data in terms of metallicity using the same [Fe/H] steps for which \citet{Amard2019} isochrones are available. Next, we select single main sequence stars as those between a 5 Gyr isochrone for the bin's upper-metallicity value shifted by $\Delta M_G=-(0.376+\sigma_{G}) + A_G(d)$ and $\Delta (BP-RP)=+\sigma_{BP-RP}+A_{BP-RP}(d)$, and a 1 Gyr isochrone for the bin's lower-metallicity value shifted by $\Delta M_G=+\sigma_{M_G}$ and $\Delta (BP-RP)=-\sigma_{BP-RP}$. $\sigma_{G}$ and $\sigma_{BP-RP}$ are the typical uncertainties in the Gaia DR3 photometry at G=20 mag. $A_G(d)$ and  $A_{BP-RP}(d)$ are the average extinction in the Kepler Field at a given distance.  

\begin{figure}
	\includegraphics[trim=0mm 0mm 6mm 0mm,width=0.9\columnwidth]{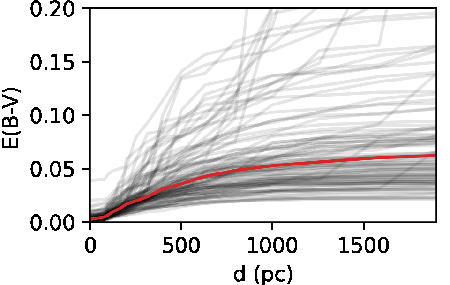}
    \caption{Extinction as a function of distance at 100 locations uniformly distributed in the Kepler field are shown in faint black. The average of these 100 extinction curves is shown in red.}
    \label{fig:ExtinctionCurve}
\end{figure}

To estimate the Kepler Field's average extinction, we applied the \verb|Bayestar17| extinction map \citep{Green2018}, which is based on Pan-STARRS 1 and 2MASS data with the \citet{Schlafly2016} extinction law. We use the \verb|Bayestar17| online tool\footnote{\url{http://argonaut.skymaps.info/}} to query 100 uniformly distributed locations within the Kepler field and retrieve data for extinction, $E(g-r)$, as a function of distance for each position. Figure \ref{fig:ExtinctionCurve} shows extinction as a function of distance as faint black lines for each of the positions queried, where we transformed the original data to $E(B-V)$ following the transformations from \citet{Green2018}. We then averaged these extinction curves as a function of distance, which is shown as a red-line in Figure \ref{fig:ExtinctionCurve}. Next, we used the average extinction curve to estimate the step in the distance required for an increase of 0.02 mag in the average extinction, and we used these to bin our dataset in terms of distance. Finally, we used the average extinction at the upper distance for each distance and metallicity bin to derive the appropriate $A_G(d)$ and  $A_{BP-RP}(d)$ for the equal-mass binaries cut.


\bsp	
\label{lastpage}
\end{document}